# Pseudo-clustering for combining data sets with multiple hierarchies

Seho Park[1] and A. James O'Malley[2,3]

1. Department of Industrial and Data Engineering, Hongik University, Seoul, South Korea

2. The Department of Biomedical Data Science, Geisel School of Medicine, Dartmouth College, Hanover, NH, U.S.A.

3. The Dartmouth Institute for Health Policy and Clinical Practice, Geisel School of Medicine, Dartmouth College, Hanover, NH, U.S.A.

Abstract

Multi-level modeling is an important approach for analyzing complex survey data using multi-stage sampling. However, estimation of multi-level models can be challenging when we combine several datasets with distinct hierarchies with sampling weights. This paper presents a method for combining multiple datasets with different hierarchical structures due to distinct informative sampling designs for the same survey. To develop an approach with complete generality, we propose to define a pseudo-cluster, a cluster containing only a singleton observation, to unify the data structure and thereby enable estimation of multi-level models incorporating sampling weights across the combined sample. We justify incorporating sampling weights at each level of the hierarchical model and in doing-so define a pseudo-likelihood estimation procedure. Simulation studies are used to illustrate the effect of incorporating sampling designs in this challenging multi-level modeling scenario. We demonstrate in the simulation studies that considering a linear mixed model with sampling weights provides unbiased estimates of model parameters and enhances the estimation of the variance components of the random effects. The proposed method is illustrated through a novel application from the National Survey of Healthcare Organizations and Systems that sought to determine which organizational characteristics or traits, as measured in the surveys, have the strongest average relationship to the percentage of depression and anxiety diagnoses in physician practices in the US.

1. Introduction

As data structures become increasingly complex, multi-level modeling is being increasingly and widely used across many research areas. Multi-level models are characterized by a regression structure for the mean parameters and variance (and covariance) parameters at each level of the model to account for residual dependencies in the data[1]. When sample surveys are conducted using a multi-stage sampling design with unequal selection probabilities, such as stratified cluster sampling, element-level units are nested within a cluster and the clusters are nested within upper level of units. Thus, the data structure underlying multi-level models often emulates the data structure that arises under multi-stage sampling.

Estimation of multi-level models, however, can be numerically challenging, especially when the dependent variable is categorical or otherwise needs to be modeled using a nonlinear model. When the study sample was obtained using a formal sampling process with known weights, the complexities multiply. Firstly, one cannot simply form a likelihood function that accounts for the weights because the weights are not observations and so there is a sense of arbitrariness in terms of incorporating them. Second, the construction of estimators and other inferences that account for the sampling design inherit all of the complexities associated with estimating regression models in survey data.

In this paper we are interested in combining and analyzing data collected from surveying organizations in a hierarchical or tiered organizational structure, including singleton or standalone organizations. This study is motivated by the National Survey of Healthcare Organizations and Systems (NSHOS) as part of an Agency for Healthcare Research and Quality (AHRQ) funded initiative[2] to obtain information about health care systems and businesses in the United States. NSHOS data are collected by separately surveying Corporate Parents and Owner Subsidiaries, and Hospitals and Practices (referred to as a Physician Practices). In the application motivating this paper, we focus on physician practices as the experimental unit. We seek to address the specific question of which practice characteristics and traits, as measured by the NSHOS survey items, are most strongly associated with the practice-level percentage of depression and anxiety diagnoses. However, the methods generalize to numerous other analysis of the analogous or similar structure. Because owner subsidiaries are nested within corporate parents, a three-level hierarchical structure with known sampling probabilities at each level exists. The structure of the NSHOS data is unique in terms of complexity of the relationship among the three levels (Figure 1). The physician practices have mixed structure as they are either standalone ("independent") entities or are owned by a corporate parent or owner subsidiary.

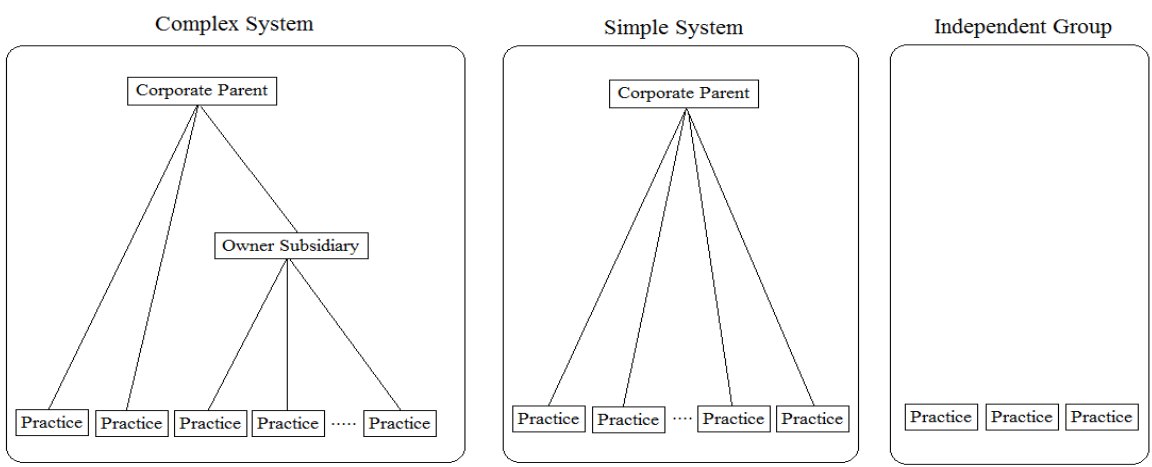

*Figure 1 Data structure of the NSHOS survey*

The physician practices in the independent group are sampled independently using a simple random sampling design. The surveys allocated to physician practices under systems were allocated according to a complex cluster-stratified design with multi-stage sampling reflecting the mix of two- (simple-system) and three-tiered (complex system) ownership structures. In complex systems physician practices are nested either directly under the corporate parent or within owner subsidiaries, which are also nested within a corporate parent. Sampling can be considered as being performed in two- or three-stages and with unequal selection probabilities. It implies that a corporate parent is considered as a primary sampling unit (PSU), an owner subsidiary is considered as a secondary sampling unit (SSU) and the physician practices are tertiary sampling units (TSU) as well as being the units of observation.

To account for informative sampling designs and thus overt getting biased results, sampling weights (the inverse of the sample selection probabilities) are incorporated in the data analysis. An alternative is to include all the design variables that determined the selection probabilities in the model as predictors. However, we are interested in developing methodology that can be applied when not all the design variables are available to analysts but the sampling weights are available. Such a scenario may arise with federal or other surveys for which some variables that defined the sampling weights are withheld to protect the privacy of the respondents. Even if all the predictors are available, it may still be the case that a model that excludes some of these variables is desired. For example, one might be interested in the average relationship between two variables in the population, not the conditional relationship given other variables that also predict the outcome.

Multi-level modeling incorporating sampling weights has been studied widely across various areas and many methods and estimators have been proposed that demonstrate justification of sampling weights in multi-level modeling. As one of the methods, a design-based probability-weighting of each level units has been proposed[3,4,5]. Pfeffermann et al.[6] propose a hierarchical model for sample as a function of both a model for population and sampling weights at each level of data and the model uses a Bayesian method for estimation. Besides, Rao et al.[7] and Yi et al.[8] suggest a design-based weighted composite likelihood approach for two-level model models that provides a design-model consistent estimator of the model parameters. Rozi et al.[9] introduce a utility of three-level model with a binary outcome in the estimation and Li and Hedeker[10] extended a two-level mixed-effect model to a three-level mixed effect model that allows covariates to influence the variances at each level.

The analytic targets in the NSHOS study include estimated regression models whose coefficients represent the effects of various predictors on outcomes, all variables measured by the survey. Because of privacy constraints, a key question is whether we are able to use the sampling weights to recover the valid (unbiased) estimates of the average effect of a predictor on the dependent variable. That is, even when the predictor of interest interacts with one of the unavailable predictors that informed the sampling designs, we hope to recover the average effect of that predictor on the outcome for the finite population represented by the sampling frame or the super-population from which the sampling frame arose. To do so we seek to incorporate sampling weights so that the estimated model is representative of the population.

Another research question is that how to combine multiple survey datasets that have different hierarchical structures to support analyses involving all units in the population. Since the NSHOS data set can be viewed as a combination of different sampling designs, it is necessary to have one unified data structure in order to use currently available statistical packages for analysis. In order to unify the survey sampling design across the distinct sampling designs, we treat each physician practice in the independent group as nested within a pseudo-cluster owner subsidiary and corporate parent of size 1 (i.e., just containing the physician practice itself). We similarly define pseudo-cluster owner-subsidiaries for physician practices that are directly under corporate parents of complex systems or in simple systems.

Considering pseudo-clusters may bring an issue of sparsity in clusters, that is the number of clusters with singletons increases drastically as number of physician practices in the independent group is considerably large. It has been recognized that the sample size at each level is an important issue in multi-level modeling and many studies examine the impact of small clusters when estimating multi-level models and a minimum of 30 units at each level is often recommended[11,12]. Hox[13] recommended a minimum of 20 observations at level-1 within 50 groups of level-2 for analysis of interactions across levels. Despite the suggested guidelines, in real data, it is inevitable to have clusters with a small number of units at each level and the recommended sample sizes are difficult or impossible to achieve. Mok[14] found that the variance estimates are biased in balanced designs with as few as five level-2 units and Clarke and Wheaton[15] found that the proportion of singletons in particular, but sample-size in general, impacts the bias in the intercept and slope variance estimates. Furthermore, Bell et al.[16,17] found that the accuracy of the confidence intervals for level-2 predictors reduced as the proportion of singletons increased. This raises concerns for our analysis given that there are a large number of singletons: organizations that can be thought of as having a cluster size of 1. Thus, one pertinent research question studied herein is the impact of having many extremely small clusters. Therefore, the relationship of cluster size in multi-level modeling to the bias and precision of our results is of particular interest.

As one possible way to estimate model parameters from combined datasets, Lee et al.[18] consider a class of weighted linear combinations of the estimating function. The source-specific estimating functions are weighted in order to form an overall estimating function for deriving the estimator for the combined data. However, the sampling weights are not incorporated into the estimating equation and they treat the two cohorts with distinct data structures as though they originated from the same target population. We suggest a method that can take account for the distinct data structures that are due to different sampling designs, with multiple (>2) levels.

In this study, we propose to use pseudo-clustering to combine singletons at level 2 and 3 with genuine three-level model data at observational unit level. We will examine 1) pseudo-clustering effects on the estimation of fixed effects and variance components estimation and 2) the role of sampling weights for modeling with unequal selection probabilities. In Section 2, we develop notation and the 3-level model that provides the motivation for the paper along with the heterogeneous data structure. In Section 3, we develop a pseudo-maximum-likelihood estimation approach for weighted estimation and demonstrate the benefits of using sampling weights in multi-level model estimation theoretically. Details of the theoretical derivation are reported in the Appendix. In Section 4, a limited simulation study is provided to investigate properties of the pseudo-maximum-likelihood approach, particularly with respect to the prevalence of singleton clusters and sample-sizes. In Section 5, empirical results of the NSHOS study are presented. Finally, Section 6 concludes the paper.

2. Methodology

2.1 Notation and Model

We develop a general notation that will be used throughout this paper. Let level-1 unit $i$ be selected using a specified sampling design with size $n_{jk}$ ($i = 1, \ldots, n_{jk}$), which is nested within the level-2 unit $j$ with size of $m_k$ ($j = 1, \ldots, m_k$), which in turn is nested within the level-3 unit $k$ with size of $l$ ($k = 1, \cdots, l$). We assume that the model matches the design hierarchy in this study. Let $y_{ijk}$ be the response variable for level-1 unit $i$ in level-2 unit $j$ and level-3 unit $k$, $x_{ijk}$ be the associated covariate vector at level-1, $x_{jk}$ be the associated covariate at level-2, and $x_k$ be

the associated covariate at level-3. In the case of a continuous-valued outcome, a base three-level model is

$$y_{ijk} = \beta_{0jk} + \beta_{1jk} x_{ijk} + e_{ijk} \quad (1)$$

$$\beta_{0jk} = \alpha_{0k} + \alpha_{01} x_{jk} + u_{0jk}$$

$$\beta_{1jk} = \alpha_{1k} + \alpha_{11} x_{jk} + u_{1jk}$$

$$\alpha_{0k} = \gamma_{00} + \gamma_{01} x_k + \tau_{0k}$$

$$\alpha_{1k} = \gamma_{10} + \gamma_{11} x_k + \tau_{1k}$$

where $e_{ijk} \sim N(0, \sigma_e^2)$ are independent and identically distributed (iid) residuals at level-1, $\mathbf{u}_{jk} = (u_{0jk}, u_{1jk})' \sim MVN(\mathbf{0}, \sigma_u^2 I_2)$ are iid random effects at level-2, $\boldsymbol{\tau}_k = (\tau_{0k}, \tau_{1k}) \sim MVN(\mathbf{0}, \sigma_\tau^2 I_3)$ are iid random effects at level-3, and they are all independent. This is a 3-level variance component model with a random slope at level-1 and random intercepts at all levels. The level-1 units within the same cluster (level-2 unit) tend to be more similar to each other and share the same variance ($\sigma_u^2$), and similarly the level-2 units within the same supercluster (level-3 unit) share the same variance ($\sigma_\tau^2$). The covariance structure of the response vector $\mathbf{y} = (y_{111}, y_{112}, \ldots, y_{l,m_l,n_{lm}})$ is a block-diagonal covariance matrix given by

$$Var(\mathbf{y}) = \begin{bmatrix} G_{(n_1)} + \sigma_\tau^2 J_{(n_1)} & & \\ & \ddots & \\ & & G_{(n_l)} + \sigma_\tau^2 J_{(n_l)} \end{bmatrix}$$

where $G_{(n_k)} = \sigma_e^2 I_{(n_k)} + \sigma_u^2 J_{(n_k)}$ is the $k^{th}$ block-diagonal matrix with components of $I_{(n)}$, a $n \times n$ identity matrix, and $J_{(n)}$, a $n \times n$ matrix of 1s. The off-diagonal elements are all zero and each block is associated with each cluster (level-3 unit and level-2 unit).

## 2.2 Pseudo-clustering

We wish to combine a three-level (complex system; dataset 1), two-level (simple system; dataset 2) and single-level (independent group; dataset 3) into a single model. To do so we construct pseudo-clusters, a cluster of size 1, whenever a cluster is missing entirely. Genuine clusters of size 1 are indistinguishable from pseudo clusters. Physician practices beneath a single-level or two-level hierarchy are treated as nested within the ownership level directly above them. We assume that the total variance combining all levels is the same for the three data structures and denote the total variance as $\sigma^2 = \sigma_e^2 + \sigma_u^2 + \sigma_\tau^2$.

Let $\mathbf{y}_i = (y_{i1}, \ldots, y_{i,l_i}), i = 1, 2, 3$, where $l_i$ is the total number of level-1 units in dataset $i$, be a response vector of dataset $i$. Then the combined response vector $\mathbf{y}^* = (\mathbf{y}_1', \mathbf{y}_2', \mathbf{y}_3')' = (y_{11}, \ldots, y_{1,l_1}, y_{21}, \cdots, y_{2,l_2}, y_{31}, \ldots, y_{3,l_3})$ has covariance structure

$$Var(\mathbf{y}^*) = \begin{bmatrix} V_1 & & 0 \\ & V_2 & \\ 0 & & V_3 \end{bmatrix},$$

where

$$V_1 = \begin{bmatrix} G_{(n_{11})} + \sigma_\tau^2 J_{(n_{11})} & & 0 \\ & \ddots & \\ 0 & & G_{(n_{1k})} + \sigma_\tau^2 J_{(n_{1k})} \end{bmatrix}$$

$G_{(n_i)} = \sigma_e^2 I_{(n_i)} + \sigma_u^2 J_{(n_i)}$ is a $i$th block-diagonal matrix,

$$V_2 = \begin{bmatrix} \sigma_e^2 I_{(n_{21})} + \sigma_c^2 J_{(n_{21})} & & 0 \\ & \ddots & \\ 0 & & \sigma_e^2 I_{(n_{2m})} + \sigma_c^2 J_{(n_{2m})} \end{bmatrix},$$

$\sigma_c^2 = \sigma_u^2 + \sigma_\tau^2$ is a sum of the level-two and level-three unit random effects,

and $V_3 = \begin{bmatrix} \sigma^2 & \cdots & 0 \\ \vdots & \ddots & \vdots \\ 0 & \cdots & \sigma^2 \end{bmatrix} = \sigma^2 I_{(l_3)}$. This model assumes that the marginal variance of observations is the same irrespective of cluster-size. The covariance matrix $V_1$ is a $l_1 \times l_1$ block-diagonal matrix, $V_2$ is a $l_2 \times l_2$ block-diagonal matrix, and $V_3$ is a $l_3 \times l_3$ diagonal matrix of $\sigma_e^2$ since all practices in dataset 1 are independently observed. Also, the three datasets are collected independently, so the off-diagonal matrices of $Var(\mathbf{y}^*)$ is a matrix of zeros.

A priori one might be concerned that the singleton pseudo-clusters will increase the percentage of singletons in the combined dataset, leading to biased variance component estimators. However, simulation studies performed by others have shown that an increasing percentage of singletons in two-level model analysis has no effect on the point or interval estimates of model parameters when the total number of clusters is large enough[16]. However, variance component estimators for multi-level models are affected by the sparseness of the dataset when the number of clusters is small[14]. In this case, variance estimators become increasingly biased as the proportion of singleton clusters increases[17].

3. Multi-level Model Estimation

In this section, the performance of maximum-likelihood estimation of multi-level models with sampling weights is evaluated. A two-level model is first constructed in order to derive theoretical results that provide intuitive support for the empirical results presented later for three-level models.

3.1 Unweighted Estimation

To gain insight into the impact of a large number of singleton observations on the estimation of fixed and random-effect parameters, we first consider the two-level random effect model with no covariates given by

$$y_{ij} = \alpha_j + \epsilon_{ij} \quad (2)$$
$$\alpha_j = \beta_0 + u_j$$

where random residual $\epsilon_{ij} \sim N(0, \sigma_e^2)$ and random effect of clusters at level-2, $u_j \sim g(\alpha_j) = N(0, \sigma_u^2)$, are independent. The first objective is to compute the marginal likelihood by integrating over the distributions of the random effects. Given the fact that the PSUs are independently collected, the likelihood function of the two-level model is given by

$$L(\boldsymbol{\theta}|\boldsymbol{y}, \boldsymbol{\alpha}) = \prod_{j=1}^{m} \int \prod_{i=1}^{n_j} f(y_{ij}|\alpha_j) g(\alpha_j) d\alpha_j$$

$$= \left(\frac{1}{\sqrt{2\pi\sigma_e^2}}\right)^{\Sigma_j n_j} \left(\frac{1}{\sqrt{2\pi\sigma_u^2}}\right)^m \prod_{j=1}^{m} \left[ \int \exp\left\{ -\frac{1}{2\sigma_e^2}(\boldsymbol{y}_j - \alpha_j \boldsymbol{1})'(\boldsymbol{y}_j - \alpha_j \boldsymbol{1}) - \frac{(\alpha_j - \beta_0)^2}{2\sigma_u^2} \right\} d\alpha_j \right]$$

where $\mathbf{y}_j = (y_{j1}, \cdots, y_{jn_j})$, $\mathbf{1}$ is a vector of 1 size of $n_j$, $f(y_{ij}|\alpha_j)$ is a distribution of $y_{ij}$ at level-1 and $g(\alpha_j)$ is a distribution of random intercept $\alpha_j$. Evaluating the integral, the log of the marginal likelihood is found to be

$$l(\boldsymbol{\theta}|\mathbf{y}, \boldsymbol{\alpha}) \propto -\frac{1}{2}\sum_{j=1}^{m}\sum_{i=1}^{n_j}\log\sigma_e^2 - \frac{1}{2}\sum_{j=1}^{m}\log\sigma_u^2 - \frac{1}{2}\sum_{j=1}^{m}\log\left(\frac{\mathbf{1}'\mathbf{1}}{2\sigma_e^2} + \frac{1}{2\sigma_u^2}\right)$$

$$-\sum_{j=1}^{m}\frac{1}{2}(\mathbf{y}_j - \beta_0\mathbf{1})'(\sigma_e^2 I_j + \sigma_u^2 \mathbf{1}\mathbf{1}')^{-1}(\mathbf{y}_j - \beta_0\mathbf{1})$$

which is a sum of the contributions to the log-likelihood from the normal distribution of each $\alpha_j$ and the multivariate normal distribution of each $\mathbf{y}_j$. More details about this derivation is provided in Appendix 1-A. To compute the estimates of model parameters, we solve the estimating equation for $\boldsymbol{\theta} = (\beta_0, \sigma_e^2, \sigma_u^2)$:

$$U(\boldsymbol{\theta}; y_{ij}) = \begin{bmatrix} \frac{\partial l}{\partial \beta_0} \\ \frac{\partial l}{\partial \sigma_e^2} \\ \frac{\partial l}{\partial \sigma_u^2} \end{bmatrix} = \mathbf{0}.$$

Closed form expressions for estimators of the fixed effects can be determined but the estimates of variance and covariance parameters are obtained by solving the estimating equation iteratively until it converges.

*Property 1.* The estimator of the fixed effect in model (2) is

$$\hat{\beta}_0 = \frac{\sum_{j=1}^{m}\frac{\mathbf{y}_j'\mathbf{1}}{n_j\hat{\sigma}_u^2 + \hat{\sigma}_e^2}}{\sum_{j=1}^{m}\frac{n_j}{n_j\hat{\sigma}_u^2 + \hat{\sigma}_e^2}} = \frac{\sum_{j=1}^{m}\frac{\bar{y}_{\cdot j}}{\hat{\sigma}_u^2 + \hat{\sigma}_e^2/n_j}}{\sum_{j=1}^{m}\frac{1}{\hat{\sigma}_u^2 + \hat{\sigma}_e^2/n_j}}$$

where the estimates of the variances, $\hat{\sigma}_u^2$ and $\hat{\sigma}_e^2$, are obtained by solving the estimating equation iteratively. Assuming clusters are collected independently, the variance of $\hat{\beta}_0$ is estimated as

$$Var(\hat{\beta}_0) = \sum_{j}^{m} w_j^2 \, Var(\bar{y}_{\cdot j}) = \sum_{j}^{m} w_j^2 \frac{\sigma_e^2}{n_j},$$

where $w_j = \frac{1/(\hat{\sigma}_u^2 + \hat{\sigma}_e^2/n_j)}{\sum_{j=1}^{m} 1/(\hat{\sigma}_u^2 + \hat{\sigma}_e^2/n_j)}$, and a robust covariance estimator of $\hat{\beta}_0$

$$\widehat{Var}(\hat{\beta}_0) = \frac{\frac{m}{m-1}\sum_{j=1}^{m}\left\{\frac{\hat{\sigma}_u^2}{\hat{\sigma}_u^2 + \hat{\sigma}_e^2/n_j}(\bar{y}_{\cdot j} - \hat{\beta}_0) - \frac{1}{m}\sum_{j=1}^{m}\frac{\hat{\sigma}_u^2}{\hat{\sigma}_u^2 + \hat{\sigma}_e^2/n_j}(\bar{y}_{\cdot j} - \hat{\beta}_0)\right\}^2}{\left(\sum_{j=1}^{m}\frac{\hat{\sigma}_u^2}{\hat{\sigma}_u^2 + \hat{\sigma}_e^2/n_j}\right)^2},$$

obtains from the sandwich formula. Other components of the variance estimators (the elements on the diagonal of the above matrix) are provided in Appendix 1-A.

If $n_j = 1$ for all $j = 1, \cdots, m$, then $\hat{\beta}_0 = \frac{1}{m}\sum_{j=1}^{m}\bar{y}_{\cdot j} = \bar{y}_{\cdot\cdot}$, which is an overall mean of $y$, that is equivalent to the maximum likelihood estimator provided by Rabe-Hesketh and Skrondal[19] for balanced data. Also, in the case where all clusters are of size 1, the variance estimator reduces to

$$\widehat{Var}(\hat{\beta}_0) = \frac{\frac{m}{m-1}\sum_{j=1}^{m}\left\{\frac{\hat{\sigma}_u^2}{\hat{\sigma}_u^2 + \hat{\sigma}_e^2}(\bar{y}_{\cdot j} - \hat{\beta}_0) - \frac{1}{m}\sum_{j=1}^{m}\frac{\hat{\sigma}_u^2}{\hat{\sigma}_u^2 + \hat{\sigma}_e^2}(\bar{y}_{\cdot j} - \hat{\beta}_0)\right\}^2}{\left(\sum_{j=1}^{m}\frac{\hat{\sigma}_u^2}{\hat{\sigma}_u^2 + \hat{\sigma}_e^2}\right)^2}$$

$$= \frac{1}{m(m-1)}\sum_{j=1}^{m}(\bar{y}_{\cdot j} - \bar{y}_{\cdot\cdot})^2$$

which is equivalent to, $\hat{\sigma}^2/n$, the variance estimator for the sample mean under simple random sampling.

3.2 Weighted Estimation: Pseudo-maximum-likelihood estimation

Complex survey designs produce unequal selection probabilities at each level of sampling and the inverse of the selection probabilities, the sampling weights, are incorporated into maximum likelihood estimation for multi-level modeling for complex survey data to define pseudo-maximum likelihood estimation[20,21,22,23].

Pseudo-likelihood estimation is widely used for bias correction when the sample used for estimation is collected using an informative sampling design. The pseudo-likelihood for multi-level models is not as straightforward as it is for single-level models because the log-likelihood function for the sample units cannot be expressed as a simple weighted sum of the sample log-likelihood[19]. Instead it is a sum across all levels of the data hierarchy; see Rabe-Hesketh and

Skrondal[19] for an example of a pseudo-likelihood of generalized linear mixed models with weights. In general, weights are added to the likelihood function by raising the relevant likelihood-function contribution to the likelihood function so that the log-pseudo-likelihood function incorporates the weights as multiplicative constants across all levels, as we now illustrate.

Under the two-level random effect model in (1) and given the fact that the PSUs are collected independently, the weighted pseudo-likelihood for a two-level model in the presence of sampling weights is given as

$$L_w(\boldsymbol{\theta}_w|\boldsymbol{y},\boldsymbol{\alpha}) = \prod_{j=1}^{m}\left[\int \prod_{i=1}^{n_j} f(y_{ij}|\alpha_j)^{w_{i|j}} g(\alpha_j) d\alpha_j\right]^{w_j}$$

$$= \prod_{j=1}^{m}\left[\int \prod_{i=1}^{n_j}\left\{\frac{1}{\sqrt{2\pi\sigma_e^2}}\exp\left(-\frac{(y_{ij}-\alpha_j)^2}{2\sigma_e^2}\right)\right\}^{w_{i|j}} \frac{1}{\sqrt{2\pi\sigma_u^2}}\exp\left(-\frac{(\alpha_j-\beta_0)^2}{2\sigma_u^2}\right) d\alpha_j\right]^{w_j}$$

$$= \left(\frac{1}{\sqrt{2\pi\sigma_e^2}}\right)^{\Sigma_j\Sigma_i w_{i|j}w_j}\left(\frac{1}{\sqrt{2\pi\sigma_u^2}}\right)^{\Sigma_j w_j}$$

$$\times \prod_{j=1}^{m}\left[\int \exp\left\{-\frac{1}{2\sigma_e^2}(\boldsymbol{y}_j-\alpha_j\boldsymbol{1})'\boldsymbol{W}_j(\boldsymbol{y}_j-\alpha_j\boldsymbol{1}) - \frac{(\alpha_j-\beta_0)^2}{2\sigma_u^2}\right\}d\alpha_j\right]^{w_j}$$

where $w_j$ is the sampling weight for cluster $j$, $w_{i|j}$ is the conditional sampling weight for individual $i$ given the selection of cluster $j$, and $\boldsymbol{W}_j = diag[\{w_{i|j}\}]$ is a diagonal $n_j \times n_j$ matrix of conditional sampling weights at level-1.

The pseudo-likelihood can be expressed as the product of two normal distributions; see Appendix 1-B for derivation. Therefore, the log of the marginal pseudo-likelihood is

$$l_w(\boldsymbol{\theta}_w|\boldsymbol{y},\boldsymbol{\alpha}) \propto -\frac{1}{2}\sum_{j=1}^{m}\sum_{i=1}^{n_j} w_{i|j}w_j \log\sigma_e^2 - \frac{1}{2}\sum_{j=1}^{m} w_j \log\sigma_u^2 - \frac{1}{2}\sum_{j=1}^{m} w_j \log\left(\frac{\boldsymbol{1}'\boldsymbol{W}_j\boldsymbol{1}}{2\sigma_e^2} + \frac{1}{2\sigma_u^2}\right)$$

$$-\frac{1}{2}\sum_{j=1}^{m} w_j[\boldsymbol{W}_j(\boldsymbol{y}_j-\beta_0\boldsymbol{1})]'\left[\sigma_e^2\boldsymbol{W}_j + \sigma_u^2(\boldsymbol{W}_j\boldsymbol{1})(\boldsymbol{W}_j\boldsymbol{1})'\right]^{-1}[\boldsymbol{W}_j(\boldsymbol{y}_j-\beta_0\boldsymbol{1})]$$

Estimates $\boldsymbol{\theta}_w = (\beta_0^*, \sigma_e^{2*}, \sigma_u^{2*})'$ are obtained by solving the estimating equations:

$$U(\boldsymbol{\theta_w}; y_{ij}) = \begin{bmatrix} \frac{\partial l_w}{\partial \beta_0} \\ \frac{\partial l_w}{\partial \sigma_e^2} \\ \frac{\partial l_w}{\partial \sigma_u^2} \end{bmatrix} = 0.$$

*Property 2*. The weighted estimator of the fixed effect in model (2) is

$$\hat{\beta}_0^* = \frac{\sum_{j=1}^m w_j \frac{y_j' W_j 1}{\hat{\sigma}_u^{2*} 1' W_j 1 + \hat{\sigma}_e^{2*}}}{\sum_{j=1}^m w_j \frac{1' W_j 1}{\hat{\sigma}_u^{2*} 1' W_j 1 + \hat{\sigma}_e^{2*}}}$$

where the estimated variances, $\hat{\sigma}_u^{2*}$ and $\hat{\sigma}_e^{2*}$, are obtained by solving the estimating equation iteratively. Assuming clusters are collected independently, a robust covariance estimator of $\hat{\beta}_0^*$ is

$$\widehat{Var}(\hat{\beta}_0^*) = \frac{\frac{m}{m-1} \sum_{j=1}^m \left\{ w_j \frac{\hat{\sigma}_u^{2*}(y' W_j 1 - 1' W_j 1 \hat{\beta}_0^*)}{\hat{\sigma}_u^{2*} 1' W_j 1 + \hat{\sigma}_e^{2*}} - \frac{1}{m} \sum_{j=1}^m w_j \frac{\hat{\sigma}_u^{2*}(y' W_j 1 - 1' W_j 1 \hat{\beta}_0^*)}{\hat{\sigma}_u^{2*} 1' W_j 1 + \hat{\sigma}_e^{2*}} \right\}^2}{\left( \sum_{j=1}^m w_j \frac{\hat{\sigma}_u^{2*} 1' W_j 1}{\hat{\sigma}_u^{2*} 1' W_j 1 + \hat{\sigma}_e^{2*}} \right)^2}$$

using the sandwich formula given by

$$\widehat{Var}(\hat{\beta}_0^*) = I(\hat{\beta}_0^*)^{-1} \widehat{Var}(\hat{U}(\hat{\beta}_0^*)) I(\hat{\beta}_0^*)^{-1}$$

where $U(\beta_0^*)$ is a score function of $\beta_0^*$. Each component of the variance estimator can be found in Appendix 1-B.

If all $n_j = 1$, then all $w_{i|j} = 1$ and the estimator becomes

$$\hat{\beta}_0^* = \frac{\sum_{j=1}^m w_j \frac{\bar{y}_{\cdot j}}{\hat{\sigma}_u^{2*} + \hat{\sigma}_e^{2*}/n_j}}{\sum_{j=1}^m w_j \frac{1}{\hat{\sigma}_u^{2*} + \hat{\sigma}_e^{2*}/n_j}} = \frac{\sum_{j=1}^m w_j \bar{y}_{\cdot j}}{\sum_{j=1}^m w_j}$$

and its variance estimator is given by

$$Var(\hat{\beta}_0^*) = \sum_j^m w_j^{*2} Var(\bar{y}_{\cdot j}) = \sum_j^m w_j^{*2} \frac{\sigma_e^2}{n_j},$$

where $w_j^* = \frac{w_j}{\Sigma_j w_j}$. The above estimator is equivalent to the unweighted maximum likelihood estimator of $\beta_0$ in (1) when all $w_j$ are the same constant for $j$, as under simple random sampling at the level of the cluster.

3.3 Three-level model

We now extend weighted two-level model estimation to weighted three-level model estimation. We use the notation introduced in Section 2 and denote the random effect at levels two and three by $\alpha_{jk}$ and $\tau_k$, respectively, and a vector of the random effects by $\boldsymbol{\alpha}$ and $\boldsymbol{\tau}$. The marginal likelihood of a vector of parameters $\boldsymbol{\theta}_3 = (\boldsymbol{\beta}, \sigma_e^2, \sigma_u^2, \sigma_\tau^2)$ is

$$L_3(\boldsymbol{\theta}_3|\boldsymbol{y},\boldsymbol{\alpha},\boldsymbol{\tau}) = \prod_{k=1}^l \int \prod_{j=1}^{m_k} \left\{ \int \prod_{i=1}^{n_{jk}} f(y_{ijk}|\alpha_{jk},\tau_k) g(\alpha_{jk}|\tau_k) d\alpha_{jk} \right\} h(\tau_k) d\tau_k$$

where $f(y_{ijk}|\alpha_{jk},\tau_k) \sim N(x_{ijk}\boldsymbol{\beta}, \sigma_e^2)$, $g(\alpha_{jk}|\tau_k) \sim N(\tau_k, \sigma_u^2)$ and $h(\tau_k) \sim N(0, \sigma_\tau^2)$. When we incorporate sampling weights in the model parameter estimation, the pseudo-maximum likelihood of the vector of parameters $\boldsymbol{\theta}_{w,3} = (\boldsymbol{\beta}, \sigma_e^2, \sigma_u^2, \sigma_\tau^2)$ is given by

$$L_{w,3}(\boldsymbol{\theta}_{w,3}|\boldsymbol{y},\boldsymbol{\alpha},\boldsymbol{\tau}) = \prod_{k=1}^l \left[ \int \prod_{j=1}^{m_k} \left\{ \int \prod_{i=1}^{n_{jk}} f(y_{ijk}|\alpha_{jk},\tau_k)^{w_{i|jk}} g(\alpha_{jk}|\tau_k) d\alpha_{jk} \right\}^{w_{j|k}} h(\tau_k) d\tau_k \right]^{w_k}$$

where $(w_k, w_{j|k}, w_{i|jk})$ are sampling weights at level-three, level-two, and level-one, respectively. Recall that the sampling weights are the inverses of the selection probabilities at the corresponding level.

3.4 Sampling weights adjustment

Estimator consistency in pseudo-maximum-likelihood estimation requires the increase of both the number of clusters and the number of units per cluster[24]. In the case of a small number of clusters, it is necessary to rescale the sampling weights to reduce the bias in variance estimation[3]. A common scaling method is to rescale the lower-level weights to sum up to actual cluster size

(method 2; Pfeffermann et al.[3]). The intuition is that the weights at the lowest level are rescaled to equal the sum of the new weights at the unit level and so on. That is, the scaled sampling weights ($w^*_{i|jk}$) at level-one for $i \in (j, k)$ cluster are

$$w^*_{i|jk} = w_{i|jk} \times \frac{n_{jk}}{\sum_i^{n_{jk}} w_{i|jk}}$$

and the scaled sampling weights ($w^*_{j|k}$) at level-two for $j \in k$ cluster are

$$w^*_{j|k} = w_{j|k} \times \frac{m_k}{\sum_j^{m_k} w_{j|k}}$$

making the sum of scaled sampling weights match the actual sample size of the upper-level cluster. In general, we posit that such re-scaling should be performed at each level under the highest level of the model.

4. Simulation

In this simulation study, we are interested in estimation of parameters in the two-level model when informative sampling designs are used at both levels. Firstly, we test the impact of the proportion of pseudo-clusters (i.e. singletons) in the sample on the estimation of the fixed effects parameters and the variances of the random effects. Second, we present the role of sampling weights in parameter estimation of the two-level model by comparing estimates of level-2 fixed effects with versus without consideration of sampling weights in parameter estimation. We consider a case when covariates related to the sampling design are not included in the fitted model; that is, the fitted model is not necessarily correct. In real survey data, the sampling weights may be correlated with variables that are not able to be released for analysis for reasons of confidentiality. In such a case, the true outcome model may depend on covariates that are not available. Through this simulation, we can determine the role of sampling weights when a misspecified model is assumed. Lastly, we demonstrate a justification of using models with random effects in the context of survey data analysis instead of non-hierarchical models when the data are hierarchically structured. We demonstrate that it is beneficial to include subject-level random effects in the model when the effects of subject-specific characteristics are of interest.

4.1 Basic setup

Let $y_{ij}$ be the response variable for level-1 unit $i$ in level-2 unit $j$, $x_{ij}$ be the associated covariate at level-1, and $(x_j, z_j)$ be the associated covariate at level-2 for $i = 1, \cdots, N_j$ and $j = 1, \cdots, M$ where $M$ is the number of clusters in the population and $N_j$ is the population size of cluster $j$. For simulation 1, we generate data for a finite population subjected to the model

$$y_{ij} = \beta_0 + \beta_1 x_{ij} + \beta_2 z_j + u_j + e_{ij} \quad (3)$$

where $(\beta_0, \beta_1, \beta_2) = (1, 1, 1)$, $x_{ij} \sim \exp(1) + 1$, $z_j \sim N(0, 1)$. For the between-cluster covariate $(z_j)$, a single value was assigned to all units in the cluster and for the within-cluster covariate $(x_{ij})$, different values were assigned to units within the cluster. The random effects $u_j$ follow $N(0, 1)$ while the random residuals $e_{ij}$ follow $N(0, 1)$ and they are independent.

For simulations 2, we generate a finite population following

$$y_{ij} = \beta_0 + \beta_1 x_j + \beta_2 z_j + \beta_3 x_j z_j + u_j + e_{ij} \quad (4)$$

where $(\beta_0, \beta_1, \beta_2, \beta_3) = (1, 1, 1, -1)$, $x_j \sim \exp(1) + 1$, $u_j \sim N(0, 0.5^2)$ and all other variables are equal to the above. In this model, an interaction term of $x_j$ and $z_j$ is included.

For all these three simulation setups, a population size of cluster $j$ is generated by $N_j = 500 \times \frac{\exp(2.5 + z_j)}{1 + \exp(2.5 + z_j)}$, ranging between 100 and 500. We first select a sample of clusters at the first-stage sampling using probability proportional to size (PPS) sampling of clusters of size $N_j$. The first-order inclusion probability of selecting cluster $j$ is defined as $p_j = m_1 \times \frac{N_j}{\Sigma_j N_j}$, where $m_1$ is a total number of clusters in the sample and $w_j = p_j^{-1}$. Once sample clusters were selected, $n_j$ units were selected by Poisson sampling with the corresponding first-order inclusion probability $p_{i|j} = n \times \frac{l_{ij}}{\Sigma_i l_{ij}}$, where $l_{ij} = 0.25$ if $e_{ij} < 0$ and $0.75$ if $e_{ij} > 0$, and $n$ is a total number of level-1 units in the sample. With this sampling design, the units with $e_{ij} > 0$ have a three times higher chance of being selected compared to the units with $e_{ij} < 0$. Similarly, the sampling weights at level-1 are given by $w_{i|j} = p_{i|j}^{-1}$. By making the sampling probability at the first-stage depend on the unobserved random effects $z_j$ and sampling probability at the second-stage depend on the residuals $e_{ij}$, the sampling design is informative at both levels.

Two samples are collected and augmented to formulate a dataset with two hierarchies: 1) One sample that has a two-level hierarchical structure is chosen from the population using cluster sampling, and 2) another sample representing singleton's group data is chosen from another population of size $N_2$ following the same distribution of the population in (3) - (5) using simple random sampling. For this sample, we treat a unit as included in the pseudo-cluster that contains only itself, and randomly choose $m_2$ units using simple random sampling. We assign a sampling weight for the pseudo-clusters of $w_j = N_2/m_2$ and a sampling weight for the level-1 unit of $w_{i|j} = 1$. The marginal variance of all observations is assumed to be the same across both samples.

We consider two different scenarios in order to test the impact of the proportion of singletons:

1) $m_2 = 0$: no singleton
2) $m_2 = 25\%$ out of $m$ clusters have singleton,

where $m$ is the total number of clusters in the sample; $m = m_1 + m_2$. When the proportion of pseudo-cluster is 0%, the sample size is a product of the number of clusters and the size of cluster, whereas, when the proportion of pseudo-cluster is 25%, the sample size decreases compared to the 0% case because 25% of the $m$ clusters are pseudo-clusters that have only one unit.

Also we consider three combinations of number of clusters and cluster size as follows: $(m, n) = (100, 30), (50, 50), (30, 100)$. The three cluster-size settings enable us to determine the impact of cluster size and the number of clusters on the estimation of parameters.

We use R for generating population data and extracting samples. We use the 'mixed' STATA function for estimating the multi-level models for continuous outcome variable.

4.2 Simulation 1

We generate data following (3) and estimate the model $y_{ij} = \beta_0 + \beta_1 x_{ij} + \beta_2 z_j + u_j + e_{ij}$ that contains the random effect term. In this case, we fit a model that is equivalent to the outcome model; i.e., as if we correctly specified the model. We repeat $B = 1,000$ times for this simulation. Results that show the bias and precision of the estimates are presented in Table 1.

4.3 Simulation 2

When we select two datasets from populations following a distribution of (4), we combine them as one dataset and fit two models as follows:

- Model 1: $y_{ij} = \beta_0 + \beta_1 x_j + u_j + e_{ij}$ that contains the random effect term to account for the variation across the clusters.
- Model 2: $y_{ij} = \beta_0 + \beta_1 x_j + e_{ij}$ that is a simple linear model and does not contain the random effect term.

We repeat the data generation and estimation $B = 1,000$ times with the results of fitting the mixed-effect and linear models presented in Table 2 and 3, respectively.

4.5 Results

In simulation 1 the model is correctly specified. As seen by the results in Table 1, the coefficients of the model parameters $(\beta_0, \beta_1, \beta_2)$ are all unbiased except for the unweighted estimates of $\beta_0$. The intercept is always biased in unweighted estimation and it is well known that the design-based estimator of the mean of a response variable, the Horvitz-Thompson estimator, is unbiased when sampling weights are incorporated in the estimation of the finite population mean.

The estimates of the upper-level variance component are unbiased under 25% setup for both weighted and unweighted estimation, whereas, under no singleton setup, unweighted estimates are biased with the amount of bias increasing as the number of clusters decreases. In terms of estimation of the residual variance, weighted estimates appear to be unbiased whereas unweighted estimates are biased under both 0% and 25% singletons. The weighted estimates consistently have little or at least less bias compared to the unweighted estimates and the bias reduces as the cluster size increases (at the same time, number of clusters decreases). As the number of level-1 units within each cluster increases (i.e. $n$ changes from 30 to 100), the variation in the level-1 units is well captured by incorporating sampling weights in estimation and gets close to the true value ($\sigma_e^2 = 1$).

In simulation 2 we misspecify the fitted model in order to examine whether applying sampling weights in estimation and using random intercept models compensates for the misspecification of the model. The results in Table 2 show that coefficient estimates ($\beta_0, \beta_1$) are unbiased in weighted estimation but biased in unweighted estimation, regardless of the proportion of pseudo-clusters, number of clusters, and cluster size. However, when the variation across cluster levels is not captured in the fitted model, such as when a simple linear regression is fitted, all weighted and unweighted estimates are biased under every setup except for the weighted estimates of $\beta_0$ that is justified by the unbiasedness property of the Horvitz-Thompson estimator. The biases of the other estimates are due to the unobserved/unmeasured variation at the cluster level, which the sampling weights cannot handle (Table 3).

The bias of the unweighted coefficient estimates in Table 2 is larger under 25% pseudo-cluster case than the 0% pseudo-cluster case. It implies that when there exist singletons in the sample, or there are some clusters having a few units within the cluster, unweighted estimation of model parameters in multi-level modeling induces considerable biases and inaccurate estimates. For the estimation of variance components in the presence of excluded covariates, the true variance of residuals is given by $\sigma_e^2 = 1$ and the true value of variance of random effect $u_j$ is

$$\begin{aligned}
\sigma_u^2 &= var(\beta_2 z_j + \beta_3 x_j z_j + u_j) \\
&= \beta_2^2 var(z_j) + \beta_3^2 var(x_j z_j) + var(u_j) + 2\beta_2\beta_3 cov(z_j, x_j z_j) \\
&= \beta_2^2 var(z_j) + \beta_3^2 E(x_j^2 z_j^2) + var(u_j) + 2\beta_2\beta_3 E(x_j z_j^2) \\
&= 1.25 + E(x_j^2) - E(x_j) \\
&= 2.25,
\end{aligned}$$

where the latter follows from the assumption that $x_j$ and $z_j$ are independent. In Table 2, we find that weighted estimates of $\sigma_u^2$ are no less biased compared to unweighted estimates especially with singletons present. The weighted estimates are closer to the true value of 2.25, and bias increases as the number of clusters decreases under both the 0% and 25% cases. Similarly, the weighted residual variance estimates provide estimates with smaller biases compared to unweighted estimations in both cases (0% and 25% singletons).

It is not reported here, but from simulation studies, we found that the coverage rates of the parameter estimates are all less than 95%. It implies that the pseudo-maximum likelihood estimation tends to produce underestimated variance estimates and it shrinkages a confidence interval of each parameter, which leads to the coverage rates that do not reach to the target coverage rate of 95%.

5. Application: NSHOS data

Goal of this empirical study is to determine the factors that influence the percentage of patients diagnosed with depression and anxiety in physician practices. A collaborative care model, or the Improving Mood Promoting Access to Collaborative Treatment (IMPACT) model, integrates mental health and primary care services. The integrated health care services are implemented by primary care practices that play an essential role in initiating depression treatment and management. Despite the benefits of collaborative care for detecting patients with depression and anxiety, the IMPACT model has been adopted by few primary care practices and the adoption ratio has not been increased from 2006 to 2013. In this study, we investigate how system, practice, and patient characteristics associate with the percentage of depression and anxiety diagnoses in physician practices using the NSHOS survey. Using beneficiary-level measures from the Medicare claims data, we merged these patient-level measurements with the NSHOS data by attributing patients to medical practices linked between the data sets.

5.1. Data Description

In the NSHOS data, there are total of 2,190 physician practices in the data: 1,482 physician practices are within systems, including complex systems and simple systems, and 708 physician practices are independent. The outcome variable is the percentage of patients diagnosed with depressions and anxiety in a physician practice. The 17 properties of physician practices and systems measured using the NSHOS are included as covariates in the model. The detailed description of the variables is attached in Appendix 2.

5.2. Data Analysis

The linear mixed-effect model we fit follows (1) with random intercepts only where $(x_k, x_{jk}, x_{ijk})$ and $(\tau_k, u_{jk}, e_{ijk})$ are covariates and random effects at corporate parent-, owner subsidiary-, and practice-level, respectively. The model includes fixed effect terms of practice-level covariates (practice size, FQHC status, geographic site, etc.), owner subsidiary-level covariate (number of owner subsidiaries, indicator of being directly under corporate parent and owner subsidiary) and system-level covariate (system type). A random intercept for each system is included to account for unmeasured system-level characteristics ($\tau_k$) or owner subsidiary-level characteristics ($u_{jk}$) that might have influence on the percentage of patients diagnosed with depression. The model is free from the unidentifiability issue that arises when all owner subsidiaries are pseudo-clusters that only one physician practice is measured within the pseudo-clusters. The unidentifiable owner subsidiaries' random intercept $u_{jk}$ are then confounded with overall error $e_{ijk}$. However, in the complex system and simple system, there are clusters with more than one unit within the clusters and it avoids the model identification problem.

5.3. Results

Results of the NSHOS data analysis are reported in Table 4 and Table 5.

Table 4 reports estimated coefficients, standard errors, z-value and p-value of fixed effects under two scenarios, weighted estimation or unweighted estimation, and Table 5 reports estimated coefficients and standard errors of variance components of two random effects under the two scenarios. Before implementing weighted estimation, sampling weights at level-1 and level-2 calibrated to fit the actual sample size of the upper level unit that correct the underestimation of the variance components[3].

Based on a significance level of 5%, region of practice, mean age of patients sex, race and poverty level appear to be significantly meaningful to predict the percentage of depression diagnoses under both weighted and unweighted estimation. That is, when the patients are young, female, Hispanic and in areas with higher poverty levels, the percentage getting diagnosed with depression is likely to be larger. Practice size and the races indicators are important covariates for explaining the relationship with the outcome variable under unweighted estimation, whereas they are not significant under weighted estimation. Because practice size is related to the selection probabilities at level-1, it contributes to the sampling weights. Accounting for survey weights may obviate the need to include such variables as predictors. Moreover, some covariates such as FQHC, system type (simple), sys_amc, and region (Northeast) have opposite signs when sampling weights are used in estimation than when not, but the magnitude of the associated coefficients are close to zero and so their impact does not appear to be not meaningful.

The estimated variance components quantify the magnitude of the unexplained variation across corporate parents and across owner subsidiaries while the estimated residual variance explains the variation within physician practices. The addition of pseudo-clusters of corporate parents and owner subsidiaries for the physician practices in the independent group leads to increased estimates of the variance components at higher levels of the model with the estimated random effect variances for corporate parents and owner subsidiaries three to five times bigger under weighted estimation compared to unweighted estimation. As noticed in the simulation study, when sampling weights are a function of unmeasured factors or latent covariate variables, excluding predictors appears to inflate variance component estimates, even more so when sampling weights are used. Incorporating sampling weights in mixed-effect models leads to the increased estimates of the variance component of random effects reduces biases in coefficient estimation. In contrast, un-weighted estimation may confound unexplained variations in the model with the residual variance and not correctly partition the amount of unexplained variation across the levels of the model.

## 6. Discussion

In this paper, we demonstrate that a multi-level model with pseudo-clustering can be employed to combine multiple datasets with different hierarchical structures and can provide unbiased estimates of model parameters incorporating sampling weights. We consider a complex sampling design at both levels, which means that we used unequal selection probabilities at each level, that generates sampling weights at both level that are informative and non-ignorable for accurate model parameter estimation. A pseudo-maximum likelihood estimator of model parameters in a linear mixed model is discussed, and simulation studies justify the role of sampling weights in estimation of a multi-level model by enhancing the estimation of model parameters in terms of accuracy regardless of the percentage of singletons. Pseudo-clustering generates artificial clusters with singletons and it leads to an increase in random variance component estimation at cluster level, but it allows the different data structures in multiple datasets to be incorporated in a single analysis.

We employ this new approach to analyze the NSHOS data that are a combination of three datasets with different structures. Pseudo-clustering of physician practices in a simple system and an independent group enables us to analyze the mix of three different hierarchical datasets using a currently available statistical package incorporating sampling weights. As demonstrated above, estimated coefficients of covariates in the model used for NSHOS data analysis under weighted estimation appear to be unbiased and captures variation between clusters that is confounded in residual variances at level-1 in unweighted estimation.

Sampling weights adjustment for units in the pseudo-clusters, for example, sampling weights adjustment for the corporate parents and owner subsidiaries in the independent group of the NSHOS data, requires caution to researchers who want to employ the adjustment method in Pfeffermann et al.[3] In such a situation, all sampling weights should become 1 in order to match the actual size of upper-level unit (pseudo-cluster), but this may violate the distribution of sampling weights assigned to the observational units. In a simulation study, we compared the scenarios for weighting physician practices in the independent group use; 1) assigned sampling weights of 1 by adopting Pfeffermann's weight adjustment, and 2) use by the sampling design weight. We found little difference and have reported the results of using the sampling weights derived by sampling design for the physician practices in the independent group in this paper. The results are encouraging for researchers who analyze multi-level model incorporating sampling weights in the presence of singletons in clusters.

We assume the same total variance for combining the three different datasets in the NSHOS into one dataset with a single level. It can be considered as a strong assumption that all datasets we try to combine should have the same amount of variance in total. This assumption can be unrealistic in practice and required to be weakened to allow heterogeneous variances for each dataset. Despite the strong restriction by assuming the same variance across the three datasets, we had to assume the homogeneous variance assumption due to the limitation in operating STATA package for the NSHOS data analysis. The STATA allows to analyze multiple datasets with heterogenous variances, but it does not support to incorporate sampling weights in multi-level modeling under the heterogeneous variance assumption.

One unique situation in our study is that multiple heterogeneous stem from heterogeneous sampling designs in datasets. The pseudo-clustering that we propose overcomes the concern when estimating multi-level model. We also justify a role of sampling weights in model parameter estimation in a multi-level models that compensate the misspecification of model fitting, diminishes biases in the model parameter estimates and recovers to unbiased estimates. Incorporating sampling weights provides more accurate estimates of variance components.

In this paper, we highlight that pseudo-clusters in multi-level modeling incorporating sampling weights are advantageous to combine multiple data sets. We also derive formula of weighted estimators of parameters in a multi-level model for continuous responses using pseudo-maximum likelihood estimation only for continuous response variable case. A theoretical demonstration of impact of incorporating sampling weights in generalized linear mixed-effect model estimation for dichotomous responses can be further studied.

Acknowledgement

This work was supported by the Agency for Healthcare Research and Quality's (AHRQ's) Comparative Health System Performance Initiative under Grant # 1U19HS024075, which studies how health care delivery systems promote evidence-based practices and patient-centered outcomes research in delivering care. The findings and conclusions in this article are those of the author(s) and do not necessarily reflect the views of AHRQ. The data that support the findings of this study are available on request from the corresponding author. The data are not publicly available due to privacy or ethical restrictions.

Tables.

*Table 1 Simulation 1 – Estimates and standard errors of coefficients and variance components under the correctly specified model fitting*

| (m, n) | % of Singletons | | 0% | | 25% | |
|---|---|---|---|---|---|---|
| | | True Value | Non-weighted | Weighted | Non-weighted | Weighted |
| (100, 30) | $\beta_0$ | 1 | 1.401 (0.118) | 1.019 (0.125) | 1.347 (0.112) | 1.018 (0.129) |
| | $\beta_1$ | 1 | 1.002 (0.034) | 1.003 (0.043) | 0.999 (0.035) | 1.000 (0.044) |
| | $\beta_2$ | 1 | 1.003 (0.105) | 1.003 (0.110) | 1.002 (0.105) | 0.997 (0.116) |
| | $\sigma_e^2$ | 1 | 0.844 (0.026) | 0.967 (0.030) | 0.844 (0.026) | 0.967 (0.030) |
| | $\sigma_u^2$ | 1 | 0.977 (0.150) | 1.010 (0.153) | 1.006 (0.144) | 1.015 (0.148) |
| (50, 50) | $\beta_0$ | 1 | 1.139 (0.157) | 1.002 (0.167) | 1.352 (0.151) | 1.017 (0.167) |
| | $\beta_1$ | 1 | 1.000 (0.038) | 1.001 (0.048) | 0.998 (0.037) | 1.000 (0.048) |
| | $\beta_2$ | 1 | 1.000 (0.155) | 0.995 (0.157) | 1.008 (0.142) | 0.998 (0.154) |
| | $\sigma_e^2$ | 1 | 0.844 (0.028) | 0.983 (0.032) | 0.843 (0.028) | 0.983 (0.032) |
| | $\sigma_u^2$ | 1 | 0.957 (0.197) | 0.963 (0.198) | 0.982 (0.200) | 0.985 (0.201) |
| (30, 100) | $\beta_0$ | 1 | 1.221 (0.198) | 1.003 (0.200) | 1.351 (0.111) | 1.010 (0.121) |
| | $\beta_1$ | 1 | 1.001 (0.044) | 1.002 (0.047) | 0.998 (0.036) | 0.999 (0.045) |
| | $\beta_2$ | 1 | 1.044 (0.203) | 1.009 (0.204) | 1.006 (0.102) | 1.000 (0.110) |
| | $\sigma_e^2$ | 1 | 0.944 (0.031) | 0.995 (0.031) | 0.844 (0.026) | 0.967 (0.030) |
| | $\sigma_u^2$ | 1 | 0.948 (0.259) | 0.947 (0.261) | 1.005 (0.141) | 1.011 (0.144) |

*Table 2 Simulation 2 - Fitted model 1 considering random effects*

| (m, n) | % of Singletons | | 0% | | 25% | |
|---|---|---|---|---|---|---|
| | | True Value | Non-weighted | Weighted | Non-weighted | Weighted |

| (m, n) | | | | | | |
|---|---|---|---|---|---|---|
| (100, 30) | $\beta_0$ | 1 | 1.480 (0.606) | 1.020 (0.642) | 1.410 (0.536) | 1.018 (0.598) |
| | $\beta_1$ | 1 | 0.913 (0.353) | 1.000 (0.375) | 0.927 (0.306) | 1.000 (0.344) |
| | $\sigma_e^2$ | 1 | 0.843 (0.025) | 0.966 (0.029) | 0.845 (0.026) | 0.966 (0.029) |
| | $\sigma_u^2$ | 2.25 | 1.935 (0.686) | 2.065 (0.763) | 2.036 (0.688) | 2.126 (0.835) |
| (50, 50) | $\beta_0$ | 1 | 1.505 (0.779) | 1.015 (0.838) | 1.426 (0.692) | 1.048 (0.790) |
| | $\beta_1$ | 1 | 0.897 (0.463) | 0.989 (0.501) | 0.919 (0.409) | 0.993 (0.472) |
| | $\sigma_e^2$ | 1 | 0.845 (0.027) | 0.984 (0.032) | 0.847 (0.029) | 0.986 (0.032) |
| | $\sigma_u^2$ | 2.25 | 1.824 (0.861) | 1.951 (1.054) | 1.931 (0.883) | 2.099 (1.105) |
| (30, 100) | $\beta_0$ | 1 | 1.300 (0.892) | 1.020 (0.943) | 1.209 (0.855) | 0.998 (0.998) |
| | $\beta_1$ | 1 | 0.902 (0.531) | 0.981 (0.565) | 0.951 (0.508) | 1.008 (0.602) |
| | $\sigma_e^2$ | 1 | 0.945 (0.030) | 0.996 (0.031) | 0.945 (0.033) | 0.997 (0.035) |
| | $\sigma_u^2$ | 2.25 | 1.616 (1.017) | 1.752 (1.169) | 1.794 (1.035) | 1.871 (1.299) |

*Table 3 Simulation 2 - Fitted model 2 using simple linear model*

| (m, n) | % of Singletons | | 0% | | 25% | |
|---|---|---|---|---|---|---|
| | | True Value | Non-weighted | Weighted | Non-weighted | Weighted |
| (100, 30) | $\beta_0$ | 1 | 1.477 (0.607) | 1.076 (0.611) | 1.462 (0.565) | 1.068 (0.575) |
| | $\beta_1$ | 1 | 0.917 (0.354) | 0.918 (0.356) | 0.919 (0.325) | 0.920 (0.331) |
| (50, 50) | $\beta_0$ | 1 | 1.509 (0.763) | 1.104 (0.783) | 1.511 (0.738) | 1.109 (0.756) |
| | $\beta_1$ | 1 | 0.888 (0.454) | 0.901 (0.467) | 0.891 (0.442) | 0.917 (0.452) |
| (30, 100) | $\beta_0$ | 1 | 1.354 (0.876) | 1.099 (0.904) | 1.299 (0.891) | 1.030 (0.935) |
| | $\beta_1$ | 1 | 0.864 (0.520) | 0.902 (0.538) | 0.894 (0.530) | 0.913 (0.558) |

*Table 4 Estimated Coefficients (Coef), Robust Standard Errors (S.E.), Z-value (z), and P-value of the Three-level Modeling on the NSHOS Data*

|  | Weighted | | | | Unweighted | | | |
|---|---|---|---|---|---|---|---|---|
| Variable | Coef | S.E. | z | p-value | Coef | S.E. | z | p-value |
| intercept | 0.433 | 0.079 | 5.500 | 0.000 | 0.466 | 0.032 | 14.610 | 0.000 |
| System type | | | | | | | | |
| Complex | -0.008 | 0.008 | -0.970 | 0.333 | -0.007 | 0.006 | -1.260 | 0.207 |
| Simple | -0.013 | 0.010 | -1.330 | 0.183 | 0.002 | 0.005 | 0.370 | 0.715 |
| Nos | 0.002 | 0.001 | 1.310 | 0.190 | 0.001 | 0.001 | 0.750 | 0.452 |
| dirOS | 0.003 | 0.007 | 0.510 | 0.608 | -0.003 | 0.006 | -0.520 | 0.605 |
| Practice Size | | | | | | | | |
| Small (4-6) | 0.008 | 0.006 | 1.320 | 0.188 | 0.005 | 0.004 | 1.230 | 0.217 |
| Medium (7-12) | 0.008 | 0.006 | 1.490 | 0.137 | 0.009 | 0.005 | 1.900 | 0.057 |
| Large (13+) | 0.009 | 0.007 | 1.320 | 0.187 | 0.011 | 0.005 | 2.180 | 0.029 |
| FQHC | -0.002 | 0.006 | -0.320 | 0.750 | 0.008 | 0.005 | 1.650 | 0.099 |
| sys_amc | -0.002 | 0.014 | -0.160 | 0.874 | 0.002 | 0.006 | 0.280 | 0.780 |
| ruca_group | -0.011 | 0.006 | -1.850 | 0.064 | -0.004 | 0.005 | -0.770 | 0.439 |
| region | | | | | | | | |
| Northeast | -0.007 | 0.009 | -0.810 | 0.418 | 0.004 | 0.005 | 0.720 | 0.475 |
| South | -0.030 | 0.009 | -3.310 | 0.001 | -0.012 | 0.005 | -2.270 | 0.023 |
| West | -0.044 | 0.015 | -2.900 | 0.004 | -0.020 | 0.006 | -3.530 | 0.000 |
| apm_par | -0.007 | 0.007 | -1.080 | 0.280 | -0.006 | 0.005 | -1.290 | 0.198 |
| any_aco_par | 0.003 | 0.005 | 0.630 | 0.531 | 0.007 | 0.004 | 1.890 | 0.059 |
| cpc_plus | | | | | | | | |
| CPC+ participant | 0.002 | 0.009 | 0.220 | 0.826 | 0.000 | 0.007 | -0.040 | 0.965 |
| Not CPC+ Eligible | -0.004 | 0.006 | -0.670 | 0.504 | -0.005 | 0.005 | -0.880 | 0.379 |
| mean_age | -0.004 | 0.001 | -3.700 | 0.000 | -0.004 | 0.000 | -10.520 | 0.000 |
| pct_female | 0.099 | 0.039 | 2.540 | 0.011 | 0.066 | 0.018 | 3.750 | 0.000 |
| pct_black | -0.006 | 0.026 | -0.230 | 0.816 | -0.052 | 0.013 | -4.120 | 0.000 |
| pct_hispanic | 0.055 | 0.027 | 2.010 | 0.044 | 0.047 | 0.013 | 3.590 | 0.000 |
| pct_other | -0.027 | 0.038 | -0.720 | 0.469 | -0.093 | 0.016 | -5.650 | 0.000 |
| poverty_level | 0.092 | 0.035 | 2.660 | 0.008 | 0.122 | 0.021 | 5.720 | 0.000 |

Table 5 Estimated Variance Component of Random Effects of the Three-level Modeling (5) on the NSHOS Data

| | Weighted | | Unweighted | |
|---|---|---|---|---|
| | Coefficient ($\times 10^2$) | Standard Error ($\times 10^2$) | Coefficient ($\times 10^2$) | Standard Error ($\times 10^2$) |
| Corporate Parent-level | 0.222 | 0.103 | 0.089 | 0.038 |
| Owner Subsidiary-level | 0.243 | 0.075 | 0.047 | 0.035 |
| Residual | 0.077 | 0.013 | 0.407 | 0.020 |

Appendices.

Appendix 1. Derivation of formula in Section 3

A. Un-weighted log-likelihood derivation

$$L(\boldsymbol{\theta}|\mathbf{y},\boldsymbol{\alpha}) = \prod_{j=1}^{m} \int \prod_{i=1}^{n_j} f(y_{ij}|\alpha_j) g(\alpha_j) d\alpha_j$$

$$= \left(\frac{1}{\sqrt{2\pi\sigma_e^2}}\right)^{\Sigma_j n_j} \left(\frac{1}{\sqrt{2\pi\sigma_u^2}}\right)^m \left(\pi \left(\frac{\mathbf{1}'\mathbf{1}}{2\sigma_e^2} + \frac{1}{2\sigma_u^2}\right)\right)^{\frac{m}{2}}$$

$$\times \prod_{j=1}^{m} \left[ \int \exp\left\{ -\left(\frac{\mathbf{1}'\mathbf{1}}{2\sigma_e^2} + \frac{1}{2\sigma_u^2}\right) \left(\alpha_j - \frac{\frac{\mathbf{y}_j'\mathbf{1}}{2\sigma_e^2} + \frac{\beta_0}{2\sigma_u^2}}{\frac{\mathbf{1}'\mathbf{1}}{2\sigma_e^2} + \frac{1}{2\sigma_u^2}}\right)^2 \right\} \exp\left\{ \frac{\left(\frac{\mathbf{y}_j'\mathbf{1}}{2\sigma_e^2} + \frac{\beta_0}{2\sigma_u^2}\right)^2}{\frac{\mathbf{1}'\mathbf{1}}{2\sigma_e^2} + \frac{1}{2\sigma_u^2}} - \frac{\mathbf{y}_j'\mathbf{y}_j}{2\sigma_e^2} \right. \right.$$

$$\left. \left. - \frac{\beta_0^2}{2\sigma_u^2} \right\} d\alpha_j \right]$$

The second exponential term can be rewritten by

$$\frac{\left(\frac{\mathbf{y}_j'\mathbf{1}}{2\sigma_e^2} + \frac{\beta_0}{2\sigma_u^2}\right)^2}{\frac{\mathbf{1}'\mathbf{1}}{2\sigma_e^2} + \frac{1}{2\sigma_u^2}} - \frac{1}{2\sigma_e^2} \mathbf{y}_j'\mathbf{y}_j - \frac{\beta_0^2}{2\sigma_u^2} = -\frac{1}{2}(\mathbf{y}_j - \beta_0\mathbf{1})'(\sigma_e^2 \mathbf{I}_j + \sigma_u^2 \mathbf{1}\mathbf{1}')^{-1}(\mathbf{y}_j - \beta_0\mathbf{1})$$

using Sherman-Morrison formula. Then, the likelihood function can be written by

$$L(\boldsymbol{\theta}|\mathbf{y},\boldsymbol{\alpha}) = \left(\frac{1}{\sqrt{2\pi\sigma_e^2}}\right)^{\Sigma_j n_j} \left(\frac{1}{\sqrt{2\pi\sigma_u^2}}\right)^m \left(\pi \left(\frac{\mathbf{1}'\mathbf{1}}{2\sigma_e^2} + \frac{1}{2\sigma_u^2}\right)\right)^{\frac{m}{2}}$$

$$\times \prod_{j=1}^{m} \int \exp\left\{ -\left(\frac{\mathbf{1}'\mathbf{1}}{2\sigma_e^2} + \frac{1}{2\sigma_u^2}\right) \left(\alpha_j - \frac{\frac{\mathbf{y}_j'\mathbf{1}}{2\sigma_e^2} + \frac{\beta_0}{2\sigma_u^2}}{\frac{\mathbf{1}'\mathbf{1}}{2\sigma_e^2} + \frac{1}{2\sigma_u^2}}\right)^2 \right\} d\alpha_j$$

$$\times \prod_{j=1}^{m} \exp\left\{ -\frac{1}{2}(\mathbf{y}_j - \beta_0\mathbf{1})'(\sigma_e^2 \mathbf{I}_j + \sigma_u^2 \mathbf{1}\mathbf{1}')^{-1}(\mathbf{y}_j - \beta_0\mathbf{1}) \right\}$$

$$\propto \int [\alpha_j|\mathbf{y}_j]d\alpha_j\,[\mathbf{y}_j]$$

where $[a]$ denotes a distribution of $a$ and $[\alpha_j|\mathbf{y}_j] \sim Normal\left(\dfrac{\frac{\mathbf{y}_j\mathbf{1}}{2\sigma_e^2}+\frac{\beta_0}{2\sigma_u^2}}{\frac{\mathbf{1}'\mathbf{1}}{2\sigma_e^2}+\frac{1}{2\sigma_u^2}},\left(\dfrac{\mathbf{1}'\mathbf{1}}{2\sigma_e^2}+\dfrac{1}{2\sigma_u^2}\right)^{-1}\right)$ and

$[\mathbf{y}_j] \sim MVN(\beta_0\mathbf{1},\ \sigma_e^2 \mathbf{I}_j + \sigma_u^2 \mathbf{1}\mathbf{1}')$.

The log of the marginal likelihood is then

$$l(\boldsymbol{\theta}|\mathbf{y},\boldsymbol{\alpha}) \propto -\frac{1}{2}\sum_{j=1}^{m}\sum_{i=1}^{n_j}\log\sigma_e^2 - \frac{1}{2}\sum_{j=1}^{m}\log\sigma_u^2 - \frac{1}{2}\sum_{j=1}^{m}\log\left(\frac{\mathbf{1}'\mathbf{1}}{2\sigma_e^2}+\frac{1}{2\sigma_u^2}\right)$$
$$- \sum_{j=1}^{m}\frac{1}{2}(\mathbf{y}_j-\beta_0\mathbf{1})'(\sigma_e^2\mathbf{I}_j+\sigma_u^2\mathbf{1}\mathbf{1}')^{-1}(\mathbf{y}_j-\beta_0\mathbf{1}).$$

The estimating equation for $\boldsymbol{\theta} = (\beta_0, \sigma_e^2, \sigma_u^2)$ are as follows:

$$U(\boldsymbol{\theta}; y_{ij}) = \begin{bmatrix}\dfrac{\partial l}{\partial \beta_0}\\ \dfrac{\partial l}{\partial \sigma_e^2}\\ \dfrac{\partial l}{\partial \sigma_u^2}\end{bmatrix} = 0,$$

where

$$\frac{\partial l}{\partial \beta_0} = \sum_{j=1}^{m}\left[\beta_0\left\{\frac{1}{2\sigma_u^2}\left(\frac{\mathbf{1}'\mathbf{1}}{2\sigma_e^2}+\frac{1}{2\sigma_u^2}\right)^{-1}-1\right\}+\left(\frac{\mathbf{1}'\mathbf{1}}{2\sigma_e^2}+\frac{1}{2\sigma_u^2}\right)^{-1}\left(\frac{\mathbf{y}_j'\mathbf{1}}{2\sigma_e^2}\right)\right],$$

$$\frac{\partial l}{\partial \sigma_e^2} = -\sum_{j=1}^{m}\sum_{i=1}^{n_j}\sigma_e^2 + \sum_{j=1}^{m}\frac{\mathbf{1}'\mathbf{1}}{2}\left(\frac{\mathbf{1}'\mathbf{1}}{2\sigma_e^2}+\frac{1}{2\sigma_u^2}\right)^{-1} + \sum_{j=1}^{m}\mathbf{y}_j'\mathbf{y}_j + \sum_{j=1}^{m}\left(\frac{\mathbf{1}'\mathbf{1}}{2\sigma_e^2}+\frac{1}{2\sigma_u^2}\right)^{-2}$$
$$\times\left\{-2\mathbf{y}_j'\mathbf{1}\left(\frac{\mathbf{y}_j'\mathbf{1}}{2\sigma_e^2}+\frac{\beta_0}{2\sigma_u^2}\right)\left(\frac{\mathbf{1}'\mathbf{1}}{2\sigma_e^2}+\frac{1}{2\sigma_u^2}\right) + \mathbf{1}'\mathbf{1}\left(\frac{\mathbf{y}_j'\mathbf{1}}{2\sigma_e^2}+\frac{\beta_0}{2\sigma_u^2}\right)^2\right\},$$

$$\frac{\partial l}{\partial \sigma_u^2} = -\sum_{j=1}^{m}\left\{\sigma_u^2 - \frac{1}{2}\left(\frac{\mathbf{1}'\mathbf{1}}{2\sigma_e^2}+\frac{1}{2\sigma_u^2}\right)^{-1}+\frac{\beta_0^2}{2}\right\} + \sum_{j=1}^{m}\left(\frac{\mathbf{1}'\mathbf{1}}{2\sigma_e^2}+\frac{1}{2\sigma_u^2}\right)^{-2}$$
$$\times\left\{-2\beta_0\left(\frac{\mathbf{y}_j'\mathbf{1}}{2\sigma_e^2}+\frac{\beta_0}{2\sigma_u^2}\right)\left(\frac{\mathbf{1}'\mathbf{1}}{2\sigma_e^2}+\frac{1}{2\sigma_u^2}\right) + \left(\frac{\mathbf{y}_j'\mathbf{1}}{2\sigma_e^2}+\frac{\beta_0}{2\sigma_u^2}\right)^2\right\}.$$

Assuming clusters are collected independently, variance estimator of $\hat{\beta}_0$ can be obtained using the sandwich formula given by

$$\widehat{Var}(\hat{\beta}_0) = I(\hat{\beta}_0)^{-1} \widehat{Var}(\widehat{U}(\hat{\beta}_0)) I(\hat{\beta}_0)^{-1} \qquad (A1)$$

where $\widehat{U}(\beta_0)$ is a score function of $\beta_0$. Each component of the variance estimator is obtained as

$$I(\hat{\beta}_0) = -\frac{\partial \widehat{U}(\beta_0)}{\partial \beta_0}\bigg|_{\beta_0=\hat{\beta}_0}$$

$$= \sum_{j=1}^{m} \left\{ \frac{1}{2\sigma_u^2} \left( \frac{\mathbf{1}'\mathbf{1}}{2\sigma_e^2} + \frac{1}{2\sigma_u^2} \right)^{-1} - 1 \right\} = \sum_{j=1}^{m} \frac{\sigma_u^2}{\sigma_u^2 + \sigma_e^2/n_j}$$

and

$$\widehat{Var}\left(\widehat{U}(\beta_0)\right) = \frac{m}{m-1} \sum_{j=1}^{m} (\mathbf{S}_{\cdot j} - \mathbf{S}_{\cdot\cdot})'(\mathbf{S}_{\cdot j} - \mathbf{S}_{\cdot\cdot})$$

$$= \frac{m}{m-1} \sum_{j=1}^{m} \left\{ \frac{\sigma_u^2}{\sigma_u^2 + \sigma_e^2/n_j}(\bar{y}_{\cdot j} - \beta_0) - \frac{1}{m}\sum_{j=1}^{m} \frac{\sigma_u^2}{\sigma_u^2 + \sigma_e^2/n_j}(\bar{y}_{\cdot j} - \beta_0) \right\}^2$$

where $\mathbf{S}_{\cdot j} = \beta_0 \left\{ \frac{1}{2\sigma_u^2} \left( \frac{\mathbf{1}'\mathbf{1}}{2\sigma_e^2} + \frac{1}{2\sigma_u^2} \right)^{-1} - 1 \right\} + \left( \frac{\mathbf{1}'\mathbf{1}}{2\sigma_e^2} + \frac{1}{2\sigma_u^2} \right)^{-1} \left( \frac{\mathbf{y}_j'\mathbf{1}}{2\sigma_e^2} \right) = \frac{\sigma_u^2}{\sigma_u^2 + \sigma_e^2/n_j}(\bar{y}_{\cdot j} - \beta_0)$. Thus, the variance estimator of $\hat{\beta}_0$ is

$$\widehat{Var}(\hat{\beta}_0) = \frac{\frac{m}{m-1} \sum_{j=1}^{m} \left\{ \frac{\hat{\sigma}_u^2}{\hat{\sigma}_u^2 + \hat{\sigma}_e^2/n_j}(\bar{y}_{\cdot j} - \hat{\beta}_0) - \frac{1}{m}\sum_{j=1}^{m} \frac{\hat{\sigma}_u^2}{\hat{\sigma}_u^2 + \hat{\sigma}_e^2/n_j}(\bar{y}_{\cdot j} - \hat{\beta}_0) \right\}^2}{\left( \sum_{j=1}^{m} \frac{\hat{\sigma}_u^2}{\hat{\sigma}_u^2 + \hat{\sigma}_e^2/n_j} \right)^2}$$

and if all $n_j = 1$, the variance estimator is

$$\widehat{Var}(\hat{\beta}_0) = \frac{\frac{m}{m-1} \sum_{j=1}^{m} \left\{ \frac{\hat{\sigma}_u^2}{\hat{\sigma}_u^2 + \hat{\sigma}_e^2}(\bar{y}_{\cdot j} - \hat{\beta}_0) - \frac{1}{m}\sum_{j=1}^{m} \frac{\hat{\sigma}_u^2}{\hat{\sigma}_u^2 + \hat{\sigma}_e^2}(\bar{y}_{\cdot j} - \hat{\beta}_0) \right\}^2}{\left( \sum_{j=1}^{m} \frac{\hat{\sigma}_u^2}{\hat{\sigma}_u^2 + \hat{\sigma}_e^2} \right)^2}$$

$$= \frac{1}{m(m-1)} \sum_{j=1}^{m} (\bar{y}_{\cdot j} - \bar{y}_{\cdot\cdot})^2.$$

B. Weighted log-likelihood

A pseudo-likelihood function in Section 3.2 can be rewritten by

$$L_w(\boldsymbol{\theta}_w|\boldsymbol{y}, \boldsymbol{\alpha}) = \left(\frac{1}{\sqrt{2\pi\sigma_e^2}}\right)^{\Sigma_j \Sigma_i w_{i|j} w_j} \left(\frac{1}{\sqrt{2\pi\sigma_u^2}}\right)^{\Sigma_j w_j}$$

$$\times \prod_{j=1}^{m} \left[ \int \exp\left\{-\left(\frac{\mathbf{1}'\boldsymbol{W}_j \mathbf{1}}{2\sigma_e^2} + \frac{1}{2\sigma_u^2}\right)\left(\alpha_j - \frac{\frac{\boldsymbol{y}_j'\boldsymbol{W}_j\mathbf{1}}{2\sigma_e^2} + \frac{\beta_0}{2\sigma_u^2}}{\frac{\mathbf{1}'\boldsymbol{W}_j\mathbf{1}}{2\sigma_e^2} + \frac{1}{2\sigma_u^2}}\right)^2\right\} \exp\left\{\left(\frac{\mathbf{1}'\boldsymbol{W}_j\mathbf{1}}{2\sigma_e^2}\right.\right.$$

$$\left.\left. + \frac{1}{2\sigma_u^2}\right)\left(\frac{\frac{\boldsymbol{y}_j'\boldsymbol{W}_j\mathbf{1}}{2\sigma_e^2} + \frac{\beta_0}{2\sigma_u^2}}{\frac{\mathbf{1}'\boldsymbol{W}_j\mathbf{1}}{2\sigma_e^2} + \frac{1}{2\sigma_u^2}}\right)^2 - \frac{\boldsymbol{y}_j'\boldsymbol{W}_j\boldsymbol{y}_j}{2\sigma_e^2} - \frac{\beta_0^2}{2\sigma_u^2}\right\} d\alpha_j \right]^{w_j}.$$

The second exponential term can be rewritten by

$$\left(\frac{\mathbf{1}'\boldsymbol{W}_j\mathbf{1}}{2\sigma_e^2} + \frac{1}{2\sigma_u^2}\right)\left(\frac{\frac{\boldsymbol{y}_j'\boldsymbol{W}_j\mathbf{1}}{2\sigma_e^2} + \frac{\beta_0}{2\sigma_u^2}}{\frac{\mathbf{1}'\boldsymbol{W}_j\mathbf{1}}{2\sigma_e^2} + \frac{1}{2\sigma_u^2}}\right)^2 - \frac{\boldsymbol{y}_j'\boldsymbol{W}_j\boldsymbol{y}_j}{2\sigma_e^2} - \frac{\beta_0^2}{2\sigma_u^2}$$

$$= -\frac{1}{2}[\boldsymbol{W}_j(\boldsymbol{y}_j - \beta_0\mathbf{1})]'[\sigma_e^2 \boldsymbol{W}_j + \sigma_u^2(\boldsymbol{W}_j\mathbf{1})(\boldsymbol{W}_j\mathbf{1})']^{-1}[\boldsymbol{W}_j(\boldsymbol{y}_j - \beta_0\mathbf{1})]$$

using Sherman-Morrison formula. Then, the log of the marginal pseudo-likelihood is

$$l_w(\boldsymbol{\theta}_w|\boldsymbol{y}, \boldsymbol{\alpha}) \propto -\frac{1}{2}\sum_{j=1}^{m}\sum_{i=1}^{n_j} w_{i|j} w_j \log \sigma_e^2 - \frac{1}{2}\sum_{j=1}^{m} w_j \log \sigma_u^2 - \frac{1}{2}\sum_{j=1}^{m} w_j \log\left(\frac{\mathbf{1}'\boldsymbol{W}_j\mathbf{1}}{2\sigma_e^2} + \frac{1}{2\sigma_u^2}\right)$$

$$-\frac{1}{2}\sum_{j=1}^{m} w_j [\boldsymbol{W}_j(\boldsymbol{y}_j - \beta_0\mathbf{1})]'[\sigma_e^2 \boldsymbol{W}_j + \sigma_u^2(\boldsymbol{W}_j\mathbf{1})(\boldsymbol{W}_j\mathbf{1})']^{-1}[\boldsymbol{W}_j(\boldsymbol{y}_j - \beta_0\mathbf{1})]$$

Then, the estimating equations for $\boldsymbol{\theta}_w = (\beta_0^*, \sigma_e^{2*}, \sigma_u^{2*})'$ are as follows:

$$U(\boldsymbol{\theta}_w; y_{ij}) = \begin{bmatrix} \frac{\partial l_w}{\partial \beta_0} \\ \frac{\partial l_w}{\partial \sigma_e^2} \\ \frac{\partial l_w}{\partial \sigma_u^2} \end{bmatrix} = 0,$$

where

$$\frac{\partial l_w}{\partial \beta_0^*} = \sum_{j=1}^{m} w_j \left[ \beta_0 \left\{ \frac{1}{2\sigma_u^2} \left( \frac{\mathbf{1}'\mathbf{W}_j\mathbf{1}}{2\sigma_e^2} + \frac{1}{2\sigma_u^2} \right)^{-1} - 1 \right\} + \left( \frac{\mathbf{1}'\mathbf{W}_j\mathbf{1}}{2\sigma_e^2} + \frac{1}{2\sigma_u^2} \right)^{-1} \left( \frac{\mathbf{y}_j'\mathbf{W}_j\mathbf{1}}{2\sigma_e^2} \right) \right],$$

$$\frac{\partial l_w}{\partial \sigma_e^{2*}} = -\frac{1}{2\sigma_e^2} \sum_{j=1}^{m} \sum_{i=1}^{n_j} w_{i|j} w_j + \frac{1}{2\sigma_e^4} \sum_{j=1}^{m} w_j \mathbf{1}'\mathbf{W}_j\mathbf{1} \left( \frac{\mathbf{1}'\mathbf{W}_j\mathbf{1}}{2\sigma_e^2} + \frac{1}{2\sigma_u^2} \right)^{-1} + \frac{1}{2\sigma_e^4} \sum_{j=1}^{m} w_j \, \mathbf{y}_j'\mathbf{W}_j\mathbf{y}_j$$

$$+ \sum_{j=1}^{m} w_j \left( \frac{\mathbf{1}'\mathbf{W}_j\mathbf{1}}{2\sigma_e^2} + \frac{1}{2\sigma_u^2} \right)^{-2} \left\{ -\left( \frac{\mathbf{y}_j'\mathbf{W}_j\mathbf{1}}{2\sigma_e^2} + \frac{\beta_0}{2\sigma_u^2} \right) \left( \frac{\mathbf{y}_j'\mathbf{W}_j\mathbf{1}}{\sigma_e^4} \right) \left( \frac{\mathbf{1}'\mathbf{W}_j\mathbf{1}}{2\sigma_e^2} + \frac{1}{2\sigma_u^2} \right) \right.$$

$$\left. + \frac{\mathbf{1}'\mathbf{W}_j\mathbf{1}}{2\sigma_e^4} \left( \frac{\mathbf{y}_j'\mathbf{W}_j\mathbf{1}}{2\sigma_e^2} + \frac{\beta_0}{2\sigma_u^2} \right)^2 \right\},$$

$$\frac{\partial l_w}{\partial \sigma_u^{2*}} = -\frac{1}{2\sigma_u^4} \sum_{j=1}^{m} w_j \left\{ \sigma_u^2 - \frac{1}{2} \left( \frac{\mathbf{1}'\mathbf{W}_j\mathbf{1}}{2\sigma_e^2} + \frac{1}{2\sigma_u^2} \right)^{-1} + \frac{\beta_0^2}{2} \right\} + \sum_{j=1}^{m} w_j \left( \frac{\mathbf{1}'\mathbf{W}_j\mathbf{1}}{2\sigma_e^2} + \frac{1}{2\sigma_u^2} \right)^{-2}$$

$$\times \left\{ -\frac{\beta_0}{\sigma_u^4} \left( \frac{\mathbf{y}_j'\mathbf{W}_j\mathbf{1}}{2\sigma_e^2} + \frac{\beta_0}{2\sigma_u^2} \right) \left( \frac{\mathbf{1}'\mathbf{W}_j\mathbf{1}}{2\sigma_e^2} + \frac{1}{2\sigma_u^2} \right) + \frac{1}{2\sigma_u^4} \left( \frac{\mathbf{y}_j'\mathbf{W}_j\mathbf{1}}{2\sigma_e^2} + \frac{\beta_0}{2\sigma_u^2} \right)^2 \right\}.$$

Assuming clusters are collected independently, variance estimator of $\hat{\beta}_0^*$ can be obtained using the sandwich formula given by (A1) and each component of the variance estimator at $\hat{\beta}_0^*$ is obtained as

$$I(\hat{\beta}_0^*) = -\frac{\partial \widehat{U}(\beta_0)}{\partial \beta_0} \bigg|_{\beta_0 = \hat{\beta}_0^*}$$

$$= \sum_{j=1}^{m} w_j \left\{ \frac{1}{2\sigma_u^2} \left( \frac{\mathbf{1}'\mathbf{W}_j\mathbf{1}}{2\sigma_e^2} + \frac{1}{2\sigma_u^2} \right)^{-1} - 1 \right\} = \sum_{j=1}^{m} w_j \frac{\sigma_u^2 \mathbf{1}'\mathbf{W}_j\mathbf{1}}{\sigma_u^2 \mathbf{1}'\mathbf{W}_j\mathbf{1} + \sigma_e^2}$$

and

$$\widehat{Var}\left(\widehat{U}(\beta_0)\right) = \frac{m}{m-1}\sum_{j=1}^{m}(S_{\cdot j} - S_{\cdot\cdot})'(S_{\cdot j} - S_{\cdot\cdot})$$

$$= \frac{m}{m-1}\sum_{j=1}^{m}\left\{w_j\frac{\sigma_u^2(y'W_j\mathbf{1} - \mathbf{1}'W_j\mathbf{1}\beta_0)}{\sigma_u^2\mathbf{1}'W_j\mathbf{1} + \sigma_e^2}\right.$$

$$\left. - \frac{1}{m}\sum_{j=1}^{m}w_j\frac{\sigma_u^2(y'W_j\mathbf{1} - \mathbf{1}'W_j\mathbf{1}\beta_0)}{\sigma_u^2\mathbf{1}'W_j\mathbf{1} + \sigma_e^2}\right\}^2$$

where $S_{\cdot j} = w_j\left[\beta_0\left\{\frac{1}{2\sigma_u^2}\left(\frac{\mathbf{1}'W_j\mathbf{1}}{2\sigma_e^2} + \frac{1}{2\sigma_u^2}\right)^{-1} - 1\right\} + \left(\frac{\mathbf{1}'W_j\mathbf{1}}{2\sigma_e^2} + \frac{1}{2\sigma_u^2}\right)^{-1}\left(\frac{y_j'W_j\mathbf{1}}{2\sigma_e^2}\right)\right] =$

$w_j\frac{\sigma_u^2(y'W_j\mathbf{1} - \mathbf{1}'W_j\mathbf{1}\beta_0)}{\sigma_u^2\mathbf{1}'W_j\mathbf{1} + \sigma_e^2}$. Thus, the variance estimator of $\hat{\beta}_0$ is

$$\widehat{Var}(\hat{\beta}_0^*) = \frac{\frac{m}{m-1}\sum_{j=1}^{m}\left\{w_j\frac{\hat{\sigma}_u^{2*}(y'W_j\mathbf{1} - \mathbf{1}'W_j\mathbf{1}\hat{\beta}_0^*)}{\hat{\sigma}_u^{2*}\mathbf{1}'W_j\mathbf{1} + \hat{\sigma}_e^{2*}} - \frac{1}{m}\sum_{j=1}^{m}w_j\frac{\hat{\sigma}_u^{2*}(y'W_j\mathbf{1} - \mathbf{1}'W_j\mathbf{1}\hat{\beta}_0^*)}{\hat{\sigma}_u^{2*}\mathbf{1}'W_j\mathbf{1} + \hat{\sigma}_e^{2*}}\right\}^2}{\left(\sum_{j=1}^{m}w_j\frac{\hat{\sigma}_u^{2*}\,'W_j\mathbf{1}}{\hat{\sigma}_u^{2*}\mathbf{1}'W_j\mathbf{1} + \hat{\sigma}_e^{2*}}\right)^2}$$

Appendix 2. Description of variables used in data application

| Variable Name | Description |
| --- | --- |
| System type | system type: complex, simple, independent |
| Nos | number of owner subsidiaries |
| dirOS | indicator whether the physician practice is directly under owner subsidiary |
| Practice Size | mean # of primary care physicians in practice |
| FQHC | whether FQHC or FQHC Lookalike |
| sys_amc | % in system that contains an AMC |
| ruca_group | rural or urban |
| region | Region: North Central, Northeast, South, West |
| apm_par | practice participation in alternative payment and delivery models of APM |
| any_aco_par | practice participation in alternative payment and delivery models of ACO |
| cpc_plus | practice participation in alternative payment and delivery models of CPC+ participant |
| mean_age | mean age |
| pct_female | % of female |
| pct_black | % of black |
| pct_hispanic | % of hispanic |
| pct_other | % of other |
| poverty_level | level of poverty |